\documentclass[a4paper,11pt]{article}
\usepackage[hmargin=3cm,vmargin=3cm]{geometry}
\usepackage{graphicx, amssymb, textcomp, hyperref, amsmath}
\title{\bf Quasi-exact Treatment of the Relativistic Generalized Isotonic Oscillator }
\author{\bf D. Agboola\footnote{d.agboola@maths.uq.edu.au}}
\date{\it School of Mathematics and Physics, The University of Queensland, \\Brisbane QLD 4072 Australia}

\begin{document}
\maketitle
\vspace{0.5in}
\noindent {\bf Abstract}~We investigate the pseudospin symmetry case of a spin-$\frac{1}{2}$ particle governed by the generalized isotonic oscillator, by presenting quasi exact polynomial solutions to Dirac equation with pseudospin symmetry vector and scalar potentials. The resulting equation is found to be  quasi-exactly solvable owing to the existence of a hidden $sl(2)$ algebraic structure. A systematic and closed form solution to the basic equation is obtained using the Bethe ansatz method. Analytic expression for the energy is obtain and the wavefunction is derived in terms of the roots to a set of Bethe ansatz equations.

\vspace{.5in}
\section{Introduction} In a recent study \cite{CPRS08}, the nonrelativistic one-dimensional quantum system described by the nonpolynomial potential of the form
\begin{equation}\label{eq:1}
V(x)=\frac{\omega^2x^2}{2}+g_a\frac{x^2-a^2}{(x^2+a^2)^2}
\end{equation}
where $a$ is a positive parameter, was shown to be exactly solvable in the particular case of $g_a=2$ and $\omega a^2=1/2$, with a general solution 
\begin{equation}\label{eq:2}
\left\{\begin{array}{lll}
\Psi_n(x)=\frac{P_n(x)}{(2x^2+1)}e^{-x^2/2},\\\\
E_n=-\frac{3}{2}+n ,\hspace{0.2in} n=0,3,4,5,\dots\end{array}\right..
\end{equation}
 where $\Psi_n(x)$ is the wavefunction and $E_n$ is the energy and the polynomial $P_n(x)$ relates to the Hermite polynomial as follows
\begin{equation}\label{eq:3}
P_n(x)=\left\{\begin{array}{lll}
1\hspace{3in}\mbox{for}\hspace{0.1in}n=0,\\\\
H_n(x)+4nH_{n-2}(x)+4n(n-3)H_{n-4}(x)\hspace{0.35in}\mbox{for}\hspace{0.1in}n=3,4,5,\dots\end{array}\right.
\end{equation}
Also, using the supersymmetric approach \cite{FS09} and for certain values of the parameters $g_a$, $a$ and $\omega$, potential \eqref{eq:1} has been shown to be a supersymmetric partner of the harmonic oscillator potential. Moreover, in a very recent work \cite{Sesma10}, by employing the M\"obius transformation, the Schr\"odinger equation for potential \eqref{eq:1} was transformed into a confluent Heun equation and a simple and efficient algorithm to solve the problem numerically irrespective of the the values of the parameters was presented. In addition, the 3$D$ case of the potential was studied for the quasi-polynomial solutions in cases where the potential parameters satisfy certain conditions and using the asymptotic iterative method, the authors obtain numerical solutions to the problem for a more general case \cite{HSY10}. 

On the other hand, due to the restricted number of exactly solvable systems, recent attentions have been on  the systems with partially solvable spectral. Such systems are said to be Quasi-Exactly Solvable (QES). Thus a quantum mechanical system is called quasi-exactly solvable, if only a finite number of eigenvalues and corresponding eigenvectors can be obtained exactly \cite{Turbiner96}.  
An essential feature of a QES system is that having seperated the asymptotic behaviours of the system, one gets an equation for the part which can be expanded as a power series of the basic variable. This equation unlike an exactly solvable equation with two-step recursive relations, possesses a three-step recursive relations for the coefficient of the power series. The complexity of the recursive relations does not allow one to guarantee the square integrability property of the wavefunction. However, by chosing a polynomial wavefunction, one can terminate the series at a certain order and then impose a sufficient condition for the normalization. By so doing, exact solutions to the original problem can be obtained, but only for certain energies and for special values of the parameters of the problem.

Solutions to QES systems have mostly been discussed in terms of the the recursion  relations of the power series coefficients, which is mostly expressed in terms of the (generalized) Heun differential equations. Although the solutions obtained in connection with the Heun equations are exact but the procedures involved are quite ambiguous, thus expunging the closed form of the solutions. In a series of recent studies \cite{23}-\cite{27}, the Bethe Ansatz Method (BAM) has been used in obtaining the solutions to QES systems. This method did not only yield exact solutions, it also preserve the closed form representation of the solutions. For instance, the BAM has been used to obtain the solutions of QES difference equation \cite{23} and the exact polynomial solutions of general quantum non-linear optical models \cite{24,25} and recently, the method has also been used to obtain the exact solutions for a family of spin-boson systems\cite{27}.

The purpose of this work is to extend the study of the Generalized Isotonic Oscillator (GIO) to a relativistic case, with in the framework of the pseudospin symmetry Dirac equation, using the Bethe Ansatz Method (BAM). In section 2, we reduce the Dirac equation with the GIO to a QES equation. A brief discussion of the Bethe ansatz method is given in section 3 followed by the solution to the reduced equation. Section 4 presents the Lie algebraic structure of the system and then some concluding remarks are given in section 5.
\section{Dirac equation with the GIO}
The Dirac equation for a single-nucleon with mass $\mu$ moving in a spherically symmetric attractive scalar $S(r)$ and replusive vector $V(r)$ GIO, with $\hbar=c=1$ is written as \cite{28}-\cite{30}  
\begin{equation}\label{eq:4}
H\Psi{({\bf r})}=E_n\Psi({\bf r})\hspace{.2in}\mbox{where}\hspace{.1in}H=\sum_{j=1}^3\hat{\alpha}_jp_j+\hat{\beta}[\mu+S(r)]+V(r) 
\end{equation}
and $E_n$ is the relativistic energy, $\{\hat{\alpha}_j\}$ and $\hat{\beta}$ are Dirac matrices defined as 

\begin{equation}\label{eq:5}
\hat{\alpha}_j=\begin{pmatrix}
0&\hat{\sigma}_j \\
\hat{\sigma}_j&0

\end{pmatrix}\hspace{.3in}\hat{\beta}=\begin{pmatrix}
\bf{1}&0 \\
0&\bf{-1}

\end{pmatrix}  
\end{equation}
where $\hat{\sigma}_j$ is the Pauli's $2\times 2$ matrices and $\hat{\beta}$ is a $2\times 2$ unit matrix, which satisfy anti-commutation relations
\begin{equation}\label{eq:6}
\begin{array}{lrl}
\hat{\alpha}_j\hat{\alpha}_k+\hat{\alpha}_k\hat{\alpha}_j&=&2\delta_{jk}\bf{1}\\
\hat{\alpha}_j\hat{\beta}+\hat{\beta}\hat{\alpha}_j&=&0\\
{\hat{\alpha}_j}^2=\hat{\beta}^2&=&\bf{1}

\end{array}
\end{equation}
and $p_j$ is the three momentum which can be written as
$$p_j=-i\partial_j=-i\frac{\partial}{\partial x_j} \hspace{.2in} 1\leqslant j\leqslant 3.$$ 
The orbital angular momentum operators $L_{jk}$, the spinor opertaors $S_{jk}$ and the total angular momentum operators $J_{jk}$ can be defined as follows:
\begin{equation}\label{eq:7}
L_{jk}=-L_{jk}=ix_j\frac{\partial}{\partial x_k}-ix_k\frac{\partial}{\partial x_j},\hspace{.2in} S_{jk}=-S_{kj}=i\hat{\alpha}_j\hat{\alpha}_k/2,\hspace{.2in} J_{jk}=L_{jk}+S_{jk}.$$
$$L^2=\sum_{j<k}^3L^2_{jk},\hspace{.2in}S^2=\sum_{j<k}^3S^2_{jk},\hspace{.2in}J^2=\sum_{j<k}^3J^2_{jk}, \hspace{.2in} 1\leqslant j< k\leqslant 3.
\end{equation}
For a spherically symmetric potential, total angular momentum operator $J_{jk}$ and the spin-orbit operator $\hat{K}=-\hat{\beta}(J^2-L^2-S^2+1/2)$ commutate with the Dirac Hamiltonian. For a given total angular momentum $j$, the eigenvalues of $\hat{K}$ are $\kappa=\pm(j+1/2)$; $\kappa=-(j+1/2)$ for aligned spin $j=\ell+\frac{1}{2}$ and $\kappa=(j+1/2)$ for unaligned spin $j=\ell-\frac{1}{2}$. Moreover, the spin-orbital quantum number $\kappa$ is related to the orbital angular quantum number $\ell$ and the pseudo-orbital angular quantum number $\tilde{\ell}=\ell+1$ by the expressions $\kappa(\kappa+1)=\ell(\ell+1)$ and $\kappa(\kappa-1)=\tilde{\ell}(\tilde{\ell}+1)$ respectively for $\kappa=\pm 1, \pm 2,\dots$. The spinor wave functions can be classified according to the  radial quantum number $n$ and the spin-orbital quantum number $\kappa$ and can be written using the Dirac-Pauli representation
\begin{equation}\label{eq:8}
\Psi({\bf r})=\frac{1}{r}\left(\begin{array}{lll}
F(r)Y_{jm}^\ell\left(\theta,\phi\right)\\\\
iG(r)Y^{\tilde{\ell}}_{jm}\left(\theta,\phi\right)
\end{array}\right)
\end{equation}
where $F(r)$ and $G(r)$ are the radial wave function of the upper- and the lower-spinor components respectively, $Y_{jm}^\ell\left(\theta,\phi\right)$ and $Y^{\tilde{\ell}}_{jm}\left(\theta,\phi\right)$ are the spherical harmonic functions coupled with the total angular momentum $j$. The orbital and the pseudo-orbital angular momentum quantum numbers for spin symmetry  $\ell$ and  and pseudospin symmetry $\tilde{\ell}$ refer to the upper- and lower-component respectively.  

Substituting Eq.\,\eqref{eq:8} into Eq.\,\eqref{eq:8}, and seperating the variables we obtain the following coupled radial Dirac equation for the spinor components:
\begin{equation}\label{eq:9}
\left(\frac{d}{dr}+\frac{\kappa}{r}\right)F(r)=[\mu+E_n-\Delta(r)]G(r)$$
$$\left(\frac{d}{dr}-\frac{\kappa}{r}\right)G(r)=[\mu-E_n+\Sigma(r)]F(r)
\end{equation}
where $\Sigma(r)=V(r)+S(r)$ and $\Delta(r)=V(r)-S(r)$. Using Eq.\,(9a) as the upper component and substituting into Eq.\,(9b), we obtain the following second order differential equations
\begin{equation}\label{eq:10}
\small\left[\frac{d^2}{dr^2}-\frac{\kappa(\kappa+1)}{r^2}-[\mu+E_n-\Delta(r)][\mu-E_n+\Sigma(r)]+\frac{\frac{d\Delta(r)}{dr}\left(\frac{d}{dr}+\frac{\kappa}{r}\right)}{[\mu+E_n-\Delta(r)]}\right]F(r)=0,$$
$$\small\left[\frac{d^2}{dr^2}-\frac{\kappa(\kappa-1)}{r^2}-[\mu+E_n-\Delta(r)][\mu-E_n+\Sigma(r)]-\frac{\frac{d\Sigma(r)}{dr}\left(\frac{d}{dr}-\frac{\kappa}{r}\right)}{[\mu-E_n+\Sigma(r)]}\right]G(r)=0. 
\end{equation} 
To solve these equations, we employ the concept of pseudospin symmetry \cite{29,31,32} in which $V(r)+S(r)=C_{ps}$, $C_{ps}$ being the pseudospin constant. This implies $\frac{d\Sigma_q(r)}{dr}=0$ and hence Eq.\,(10b) takes a simple form
\begin{equation}\label{eq:11}
\left\{\frac{d^2}{dr^2}-\frac{\kappa(\kappa-1)}{r^2}-\left[\mu+E_n-\Delta(r)\right]\left[\mu-E_n+C_{ps}\right]\right\}G(r)=0. 
\end{equation}
If we take 
\begin{equation}\label{eq:12}
\Delta(r)=\omega^2r^2+2g\frac{r^2-a^2}{(r^2+a^2)^2},
\end{equation}
and introduce the dimensionless quantity $z=\beta_n\omega r^2$, Eq.\,(11) becomes
\begin{equation}\label{eq:13}
zG''(z)+\frac{1}{2}G'(z)-\left[\alpha+\frac{z}{4}+\frac{\kappa(\kappa-1)}{4z}+\frac{g\left(z-\omega a^2\beta_n\right)}{2\left(z+\omega a^2\beta_n\right)^2}\right]G(z)=0.
\end{equation}
where 
\begin{equation}\label{eq:14}
-\beta^2_n=\mu-E_n+C_{ps}\hspace{.2in}\mbox{and}\hspace{0.2in}\alpha=-\frac{\left(\beta^2_n+2\mu+C_{ps}\right)\beta_n}{4\omega}
\end{equation}
From the asymptotic behaviour of Eq.\,\eqref{eq:13}, one may seek a solution of the form 
\begin{equation}\label{eq:15}
G(z)=(z+\beta_n\omega  a^2)^{b+1}~z^{{\kappa}/{2}}~e^{-z/2}~\psi(z),
\end{equation} to obtain
\begin{equation}\label{eq:16}
zG ''(z)+\left[\left(\frac{5}{2}+2b+\kappa\right)-\frac{2(b+1)\beta_n\omega a^2}{z+\beta_n\omega a^2}-z\right]G '(z)\hspace{.6in}$$
$$\hspace{0.3in}-\left[\frac{\beta_n^2g/2-(b+1)(\kappa+b+\beta_n\omega a^2+1/2)}{z+\beta_n\omega a^2} 
\right]G(z)=\left[\alpha+b+\frac{\kappa}{2}+\frac{5}{4}\right]G(z).
\end{equation} where we have assumed
\begin{equation}\label{eq:17}
b(b+1)-\beta_n^2g=0\hspace{0.1in}\Rightarrow\hspace{0.1in}b=\frac{-1-\sqrt{1+4\beta_n^2g}}{2}.
\end{equation}
It can be checked that a further transformation
\begin{equation}\label{eq:18}
G(z)=z^\nu \varphi(t),\hspace{0.2in}t=z+\beta_n\omega a^2
\end{equation}
does not change the structure of the differential equation \eqref{eq:16} 
provided $\nu=0$ and $\nu=-\kappa+1/2$. This indicates that the solutions are of double algebraic sectors, with the even solution correponding to $\nu=0$~($\kappa>0$) and the odd solution corresponding to $\nu=-\kappa+1/2$~($\kappa<0$). Thus if we use the change in variable $t=z+\beta_n\omega a^2$, Eq.\,\eqref{eq:17} takes the form 
\begin{equation}\label{eq:20}
t(t-\beta_n\omega a^2)\varphi ''(t)+\left[\left(\frac{5}{2}+2b+\kappa+2\nu+\beta_n\omega a^2\right)t-t^2-2\beta_n\omega a^2(b+1)\right]\varphi '(t)$$
$$-\left[t\left(\alpha+b+\nu+\frac{\kappa}{2}+\frac{5}{4}\right)\right]\varphi(t)=\left[\beta_n^2g/2-(b+1)(\kappa+b+2\nu+\beta_n\omega a^2+1/2)\right]\varphi(t). 
\end{equation}
\section{The Bethe ansatz solutions to relativistic GIO}
In this section, we give a brief description of the BAM of solving QES equation and then use the results to solve Eq.\,\eqref{eq:20}. For interested reader, detailed account of the method can be found in Ref.\cite{27}. 
We consider the differential equation of the form 
\begin{equation}\label{eq:21}
\left[P(t)\frac{d^2}{dt^2}+Q(t)\frac{d}{dt}+R(t)\right]S(t)=0,
\end{equation}
where
$P(t), Q(t)$ are polynomials of degree 2 and $R(t)$ is a polynomial of degree 1, which we write as
\begin{equation}\label{eq:22}
P(t)=\sum_{k=0}^2p_kt^k,\hspace{0.2in}Q(t)=\sum_{k=0}^2q_kt^k,\hspace{0.2in}R(t)=\sum_{k=0}^1r_kt^k.
\end{equation}
where $p_k$, $q_k$ and $r_k$ are constants.
This equation is quasi-exactly solvable for 
certain values of its parameters, and exact solutions are 
given by degree $n$ polynomials in $t$ with $n$ being non-negative integers. 
In fact, this equation is a special case of the general 2nd order differential equations solved in \cite{27} by means of the BAM. Applying the results in \cite{27}, we have  \\\\
{\bf Proposition 1} {\it Given a pair of polynomials $P(t)$ and $Q(t)$, then the values of the coefficients $r_0$ and $r_1$ of polynomial $R(t)$ such that the differential equation Eq.\,\eqref{eq:21} has a degree $n$ polynomial solution 
\begin{equation}\label{eq:23}
S(t)=\prod_{i=1}^n(t-t_i),\hspace{0.1in}S(t)\equiv 1\hspace{0.1in}\mbox{for}\hspace{0.1in} n=0
\end{equation}
with distinct roots $t_1, t_2,\dots,t_n$ are given by 
%\begin{equation}\label{eq:3.4}
%-c_2=nb_3,
%\end{equation}
\begin{equation}\label{eq:24}
-r_1=nq_2,
\end{equation}
\begin{equation}\label{eq:25}
-r_0=q_2\sum_{i=1}^nt_i+n(n-1)p_2+nq_1,
\end{equation}
where the roots $t_1,t_2,\dots,t_n$ satisfy the Bethe ansatz equations
\begin{equation}\label{eq:26}
\sum_{i\neq j}^n\frac{2}{t_i-t_j}+\frac{q_2t_i^2+q_1t_i+q_0}{p_2t_i^2+p_1t_i+p_0}=0,\hspace{0.1in}i=1,2,\dots,n
\end{equation}
The above equations \eqref{eq:24}--\eqref{eq:26} give all polynomial $R(t)$ such that the ODE \eqref{eq:21} has a polynomial solution \eqref{eq:23}.}

It is interesting to note that in line with the recent work \cite{SHCY11}, Eqs.\,\eqref{eq:24}, \eqref{eq:25} and \eqref{eq:26} satisfy the necessary and sufficient conditions for the differential equation Eq.\,\eqref{eq:21} to have polynomial solutions. It is easy to show that the necessary condition (2.10) of \cite{SHCY11} reduces to Eq.\,\eqref{eq:24} for $a_{3,0}=0$ and for instance, if we consider the case $n=1$, then tridiagonal determinant (Eq.\,(2.11) of \cite{SHCY11}) for Eq.\,\eqref{eq:21} takes the form
\begin{equation}
\begin{vmatrix}
-r_0&-q_0 \\
-r_1&-r_0-q_1

\end{vmatrix}=0\hspace{0.2in}\Rightarrow\hspace{0.2in} r_0=\frac{-q_1\pm\sqrt{q_1^2-4q_0q_2}}{2}, 
\end{equation}
where we have used the necessary condition \eqref{eq:24}. This result can be easily  obtained by solving for parameter $t$ in  Bethe ansatz equation \eqref{eq:26} and substituting the value into Eq.\,\eqref{eq:25}. Thus, Eqs.\,\eqref{eq:25} and \eqref{eq:26} are the sufficient conditions for differential equation \eqref{eq:21}  to have a exact polynomial solution \eqref{eq:23}.
However, it is important to note that one of the main tasks in the application of BAM is obtaining the roots of the $n$ algebraic Bethe ansatz equations \eqref{eq:26}. For an arbitrary $n$, the equation is very difficult, if not impossible, to solve algebraically. However, numerical solutions to the Bethe ansatz equations have also been discussed in many applications \cite{33}-\cite{37}. 

%\section{Relativistic energies and wavefunctions}
By comparing Eqs.\,\eqref{eq:20} and \eqref{eq:21}, we have $p_2=1$, $p_1=-\beta_n\omega a^2$, $q_2=-1$, $q_1=\left(\frac{5}{2}+2b+\kappa+2\nu+\beta_n\omega a^2\right)$, $q_0=-2\beta_n\omega a^2(b+1)$, $r_1=-\left(\alpha+b+\nu+\frac{\kappa}{2}+\frac{5}{4}\right)$ and $r_0=-\left[\beta_n^2g/2-(b+1)(\kappa+b+2\nu+\beta_n\omega a^2+1/2)\right]$.
Thus by Eqs.\,\eqref{eq:14}, \eqref{eq:17} and \eqref{eq:24}, we immediately have the energy equation
\begin{equation}\label{eq:32}
4\omega\left(n+\nu+\frac{\kappa}{2}+\frac{3}{4}\right)=\left(\beta_n^2+2\mu+C_{ps}\right)\beta_n+2\omega\sqrt{1+4g\beta_n^2}
\end{equation}
and Eq.\,\eqref{eq:25} yields
\begin{equation}\label{eq:33}
n\left(n+2b+2\nu+\kappa+\beta_n\omega a^2+3/2\right)-\sum_{i=1}^nt_\alpha=\frac{\beta_n^2g}{2}-(b+1)(\kappa+b+2\nu+\beta_n\omega a^2+1/2),
\end{equation}
with the roots $t_1,t_2,\dots t_n$ satisfying the Bethe ansatz equation
\begin{equation}\label{eq:34}
\sum_{i\neq j}^n\frac{2}{t_i-t_j}+\frac{\left(2b+\kappa+2\nu+\beta_n\omega a^2+5/2\right)t_i-t_i^2-2\beta_n\omega a^2(b+1)}{t_i(t_i-\beta_n\omega a^2)}=0
\end{equation}for $i=1,2,\dots,n$.

It is obvious from the energy equation \eqref{eq:32} that one deals with solutions with positive energy states. Moreover, the rhs of the energy equation \eqref{eq:32} remain unchange for quantum states ($n,\kappa$) and ($n-1,\kappa+2$) thereby signifying degeneracy of the energy levels between these states. This energy degeneracy does not depend on the potential parameters as it can be seen from the numerical energy values of the ground state and some excited states for the exact pseudospin case (Table 1). And we also note that for a given state, the energy values are inveresly proportional to parameter $g$.
Moreover, for the ground state, $n=0$, we have from Eq.\,\eqref{eq:33}
\begin{equation}\label{eq:35}
\frac{\beta_0^2g}{2}-(b+1)(\kappa+b+2\nu+\beta_0\omega a^2+1/2)=0,
\end{equation}
which yields the following condition:
%\left\{\begin{array}{lll}
\begin{subequations}
\begin{equation}\label{eq:36a}
\left(4\omega^2a^4-g\right)\beta_0^2+2\omega a^2(1+4\kappa)\beta_0+2\kappa(2\kappa+1)=0\hspace{0.1in}\mbox{for}\hspace{0.1in}\nu=0\\\\
\end{equation}
\begin{equation}\label{eq:36b}
\left(4\omega^2a^4-g\right)\beta_0^2+2\omega a^2(5-4\kappa)\beta_0+(4\kappa-6)(\kappa-1)=0\hspace{0.1in}\mbox{for}\hspace{0.1in}\nu=-\kappa+1/2
\end{equation}
\end{subequations}
%\end{array}\right..
Hence the ground state wavefunctions for even sector ($\nu=0$)~($\kappa>0$) can be written as
\begin{equation}\label{eq:37}
\left(\begin{array}{lll}
F_0(r)\\\\
G_0(r)
\end{array}\right)^{even}\sim r^\kappa \left(r^2+a^2\right)^be^{-\frac{\beta_0\omega r^2}{2}}\left(\begin{array}{lll}
\frac{\beta_0\omega r^3-\left(2b-\beta_0\omega a^2+2\right)r}{\beta_0^2}\\\\
\hfil r^2+a^2
\end{array}\right),
\end{equation}
with the parameters satisfying Eq.\,\eqref{eq:36a}.
Similarly, the odd sector ($\nu=-\kappa+1/2$)~($\kappa<0$) solutions are 
\begin{equation}\label{eq:38}
\left(\begin{array}{lll}
F_0(r)\\\\
G_0(r)
\end{array}\right)^{odd}\sim r^{-\kappa} \left(r^2+a^2\right)^be^{-\frac{\beta_0\omega r^2}{2}}\left(\begin{array}{lll}
\frac{\beta_0\omega r^4-\left(2b-2\kappa-\beta_0\omega a^2+3\right)r^2+(2\kappa-1)a^2}{\beta_0^2}\\\\
\hfil r\left(r^2+a^2\right)
\end{array}\right),
\end{equation}
with the parameters satisfying Eq.\,\eqref{eq:36b} and $\beta_0$ is related to ground state energy and obtained from Eq.\,\eqref{eq:32} as
\begin{equation}\label{eq:38a}
4\omega\left(\nu+\frac{\kappa}{2}+\frac{3}{4}\right)=\left(\beta_0^2+2\mu+C_{ps}\right)\beta_0+2\omega\sqrt{1+4g\beta_0^2}.
\end{equation}
 The wave functions $F_0(r)$ and $G_0(r)$ do not have nodes and 
so the states described by them are ground states of the system. 

Similarly for $n=1$, the Bethe ansatz equation \eqref{eq:34} becomes
\begin{equation}\label{eq:39}
(2b+\kappa+2\nu+\beta_1\omega a^2+5/2)t_1-t_1^2-2\beta_1\omega a^2(b+1)=0,
\end{equation}
which yields 
$$t_1=b+\nu+\frac{\kappa}{2}+\frac{\beta_1\omega a^2}{2}+\frac{5}{4}\pm\hspace{1in}$$$$\frac{1}{2}\sqrt{2(b+\nu)(2b+2\nu+2\kappa+5)-\beta_1\omega a^2(4b+8\kappa+16\nu-\beta_1\omega a^2+3)+(\kappa+5/2)^2}.$$ 
We substitute the roots into Eq.\,\eqref{eq:33} and solve the resulting algebraic equation to obtain the following condition on the parameter
\begin{subequations}
\begin{equation}\label{eq:40a}
g\beta_1^6\left(16\omega^4a^8+g^2-8g\omega^2a^4\right)+16g\omega a^2\beta^5_1\left(4\omega^2a^4-g\right)\left(\kappa+1\right)$$$$+\beta_1^4\left[4\omega^2a^4g\left(24\kappa^2+48\kappa+11-8\omega^2 a^4\right)-2g^2(4\kappa^2+8\kappa-9)\right]$$$$+\beta_1^3\left[4g\omega a^2(16\kappa^3+48\kappa^2-38\kappa-9)-16\omega^3a^6(8\kappa+11)\right]$$$$+\beta_1^2\left[2g\left(8\kappa^4+32\kappa^3+54\kappa^2+26\kappa+3\right)-24\omega^2a^4\left(8\kappa^2+18\kappa+11\right)\right]$$$$-4\omega a^2\beta_1\left(32\kappa^3+84\kappa^2+64\kappa+15\right)-4\kappa\left(8\kappa^3+20\kappa^2+16\kappa+3\right)=0
\end{equation}
\begin{equation}\label{eq:40b}
g\beta_1^6\left(16\omega^4a^8+g^2-8g\omega^2a^4\right)+16g\omega a^2\beta_1^5\left(-4\kappa\omega^2a^4+8\omega^2a^4+\kappa g-2g\right)$$$$+2\beta^4\left[g^2\left(4\kappa^2+16\kappa-21\right)+2g\omega^2a^4\left(24\kappa^2-96\kappa+83\right)-16\omega^4a^8\right]$$$$+\beta_1^3\left[4g\omega a^2(-16\kappa^3+96\kappa^2-182\kappa+93)-16\omega^3a^6(8\kappa+19)\right]$$$$+\beta_1^2\left[4g\left(4\kappa^4-32\kappa^3+99\kappa^2-131\kappa+66\right)+8\omega^2a^4\left(-24\kappa^2+102\kappa-111\right)\right]$$$$+4\omega a^2\beta_1\left(32\kappa^3-180\kappa^2+328\kappa-195\right)+4\left(-8\kappa^4+52\kappa^3-122\kappa^2+123\kappa-45\right)=0
\end{equation}
\end{subequations}
where $\beta_1$is related to the first excited energy and its obtained from Eq.\,\eqref{eq:32}  as
\begin{equation}\label{eq:41a}
4\omega\left(\nu+\frac{\kappa}{2}+\frac{7}{4}\right)=\left(\beta_1^2+2\mu+C_{ps}\right)\beta_1+2\omega\sqrt{1+4g\beta_1^2}.
\end{equation}
Thus the wavefunctions for the first excited state for the even sector ($\nu=0$)~($\kappa>0$) can be written as 
\begin{equation}\label{eq:41}
\left(\begin{array}{lll}
F_1(r)\\\\
G_1(r)
\end{array}\right)^{even}\sim r^\kappa \left(r^2+a^2\right)^be^{-\frac{\beta_1\omega r^2}{2}}\times\hspace{1.5in}$$$$\hspace{1.5in}\left(\begin{array}{lll}
\frac{\beta_1\omega r^5-\beta_1\omega\left(2b-2\beta_1\omega a^2+t_1^e+4\right)r^3-\left[\beta_1\omega a^2(2b-\beta_1\omega a^2+t_1^e+4)-2t_1^e(b+1)\right]r}{\beta^2_1}\\\\
\hfil (r^2+a^2)\left(r^2+a^2-t_1^{e}\right)
\end{array}\right),
\end{equation}
with the parameters satisfying Eq.\,\eqref{eq:40a} and the root given as
$$t_1^e=b+\frac{\kappa}{2}+\frac{\beta_1\omega a^2}{2}+\frac{5}{4}\pm\frac{1}{2}\sqrt{2b(2b+2\kappa+5)-\beta_1\omega a^2(4b+8\kappa-\beta_1\omega a^2+3)+(\kappa+5/2)^2}.$$ 
Similarly, the odd sector ($\nu=-\kappa+1/2$)~($\kappa<0$) solutions are 
\begin{equation}\label{eq:42}
\left(\begin{array}{lll}
F_1(r)\\\\
G_1(r)
\end{array}\right)^{odd}\sim r^{-\kappa} \left(r^2+a^2\right)^be^{-\frac{\beta_1\omega r^2}{2}}\times\hspace{1.5in}$$$$\hspace{1.5in}\left(\begin{array}{lll}
\frac{\left(r^2+a^2\right)\left[\beta_1^2\omega^2r^4+\beta_1\omega\left(\beta_1\omega a^2+\kappa-t_1^o-2\right)r^2+\kappa\left(\beta_1\omega a^2-t_1^o\right)\right]-2r(b+1)}{\beta_1^2}\\\\
\hfil r(r^2+a^2)\left(r^2+a^2-t_1^{o}\right)
\end{array}\right),
\end{equation}
with the parameters satisfying Eq.\,\eqref{eq:40b} and the root given as
$$t_1^o=b-\frac{\kappa}{2}+\frac{\beta_1\omega a^2}{2}+\frac{7}{4}\pm\frac{1}{2}\sqrt{3(b+2)(2b-2\kappa+1)-\beta_1\omega a^2(4b-8\kappa-\beta_1\omega a^2+11)+(\kappa+5/2)^2}.$$ 
\begin{table}[h]
\caption{Relativistic energy values of the GIO for various $n$ and $\kappa$ and a special case of $C_{ps}=0$, $g=\left\{\frac{1}{2},~2,~4\right\}$ and $\mu=\omega=1$.}

\centering
\begin{tabular}{p{.5in}l p{.8in}c c  c c c}\hline
%\multicolumn{3}{|c|}{$E(\alpha,n,D)$}\\\hline
%\multicolumn{6}{|c|}{$C_1=V_0=S_0=\mu=1$}\\\hline
$ {n}$&$\kappa>0$&\hfil$g=\frac{1}{2}$&\hfil$g=2$&$g=4$\\\hline
&1&0.7779142&0.5545547&0.4385638\\
&2&1.0709485&0.7934184&0.6363539\\
0&3&1.3018103&1.0000000&0.8157561\\
&4&1.4942015&1.1824728&0.9806115\\
&5&1.6601782&1.3457688&1.1328206\\\hline
%5&1, -6&1$h_{9/2}$&\hfil---&\hfil---&0.0203344\\
&1&1.3018103&1.0000000&0.8157561\\
&2&1.4942015&1.1824728&0.9806115\\
1&3&1.6601782&1.3457688&1.1328206\\
&4&1.8068160&1.4934769&1.2738521\\
&5&1.9386413&1.6283325&1.4049965\\\hline
%5&2, -6&2$h_{9/2}$&\hfil---&\hfil---&0.1062866\\
&1&1.6601782&1.3457688&1.1328206\\
&2&1.8068160&1.4934769&1.2738521\\
2&3&1.9386413&1.6283325&1.4049965\\
&4&2.0587311&1.7524556&1.5274033\\
&5&2.1692751&1.8675089&1.6420843\\\hline
&1&1.9386413&1.6283325&1.4049965\\
&2&2.0587311&1.7524556&1.5274033\\
3&3&2.1692751&1.8675089&1.6420843\\
&4&2.2718886&1.9748119&1.7499209\\
&5&2.3677989&2.0754239&1.8516754\\\hline
\end{tabular}

\label{tab:}
\end{table}

\section{Hidden Lie Algebraic Structure}
One way to understand the QES theory is to demonstrate that the Hamiltonian  can be expressed in terms of generator of a Lie algebra
\begin{equation}\label{eq:43}
J^-=\frac{d}{dt},\hspace{0.2in}J^+=t^2\frac{d}{dt}-nt,\hspace{0.2in}J^0=t\frac{d}{dt}-\frac{n}{2},
\end{equation}
which are differential operator realization of the $n+1$ dimensional representation of the $sl(2)$ algebra. Moreover, if  we write the basic equation \eqref{eq:21} in the Schr\"odinger form
\begin{equation}\label{eq:44}
HS(t)=-r_0S(t)
\end{equation}
where $-r_0$ is the eigenvalue of the Hamiltonian $H$, then it can easily be shown that if $r_1=-nq_2$, with $n$ being any non-negative integer, the differential operator $H$ is an element of the enveloping algebra of Lie algebra $sl(2)$
\begin{equation}\label{eq:45}
H=J^0J^0+p_1J^0J^-+q_2J^++\left(q_1+n-1\right)J^0+\left(q_0+\frac{np_1}{2}\right)J^-+\frac{n}{2}\left(\frac{n}{2}+q_1-1\right).
\end{equation}
Thus for Eq.\,\eqref{eq:20}, we have 
\begin{equation}\label{eq:46}
H=t(t-\beta_n\omega a^2)\frac{d^2}{dt^2}+\left[\left(\frac{5}{2}+2b+\kappa+2\nu+\beta_n\omega a^2\right)t-t^2-2\beta_n\omega a^2(b+1)\right]\frac{d}{dt}$$
$$\hspace{2in}-\left[t\left(\alpha+b+\nu+\frac{\kappa}{2}+\frac{5}{4}\right)\right]
\end{equation}
and 
\begin{equation}\label{eq:47}
r_0=(b+1)(\kappa+b+2\nu+\beta_n\omega a^2+1/2)-\beta_n^2g/2
\end{equation}
with $sl(2)$ algebralization
\begin{equation}\label{eq:48}
H=J^0J^0-\beta_n\omega a^2J^0J^--J^++\left(n+2b+\kappa+2\nu+\beta_n\omega a^2+\frac{3}{2}\right)J^0$$$$-\beta_n\omega a^2\left(2b+2-\frac{n}{2}\right)J^-+\frac{n}{2}\left(2b+\kappa+2\nu+\beta_n\omega a^2+\frac{n}{2}+\frac{3}{2}\right)
\end{equation}
\section{Concluding Remarks}
In this paper, we have extended the works on the GIO to a relativistic case by constructed the Bethe ansatz solutions to GIO, within the frame work of the relativistic Dirac equation. We showed that the governing equation is reducible to a QES differential equation which has an exact solution, provided the parameters satisfy certain constraints. Unlike previous non-relativistic cases the quasi-exact solvability of the equations has enabled us to use Proposition 1 to obtain closed form expressions for the energies and eigenfunctions. It is interesting to note that with the limits $\beta_n\rightarrow 1$, $\kappa\rightarrow\ell+1$, the spinor component of the wavefunction, $G(z)$ gives the non-relativistic wavefunction of the GIO, which is in agreement with previous works \cite{CPRS08,FS09,Sesma10,HSY10,SHCY11,AZ11}

Moreover, we reported the existence of degeneracies between the energy levels and the energy is inversely related with the potential parameter $g$.
We also showed that the relativistic GIO possesses an underlying $sl(2)$ algebraic structure , which is responsible for the
quasi-exact solvability of this model. Let us remark, however, that the existence of a underlying Lie algebraic 
structure in a differential equation is only a sufficient condition for the differential equation to be 
quasi-exactly solvable. In fact there are more general (than the Lie-algebraically based)
differential equations which do not possess a underlying Lie algebraic structure but are
nevertheless quasi-exactly solvable (i.e have exact polynomial solutions ) \cite{27}.
Finally, it is pertinent to note that our method gives a more general closed form expressions for the solutions, however the determination of the roots of the Bethe ansatz equations for higher exicted states may be a major difficulty in the application of the method.   
\section*{Acknowledgements}
DA wishes to thank the referee for his useful suggestions which have improved the manuscript. He is also indebted to Father J, Agboola B and Y.-Z. Zhang for their support during the preparation of the manuscript.

\end{document}